\documentclass[12pt]{article}
\usepackage{amsmath,amssymb,amsfonts,amsthm}
\title{On the non-classicality features of new classes of nonlinear coherent states}

 \author{M. K. Tavassoly
\\
\footnotesize{Atomic and Molecular Group, Faculty  of Physics,
Yazd University, Yazd, Iran}
\\ \footnotesize{e-mail: mktavassoly@yazduni.ac.ir  } }
\begin{document}

 \date{\today}


 \newcommand{\I}{\mathbb{I}}
 \newcommand{\norm}[1]{\left\Vert#1\right\Vert}
 \newcommand{\abs}[1]{\left\vert#1\right\vert}
 \newcommand{\set}[1]{\left\{#1\right\}}
 \newcommand{\R}{\mathbb R}
 \newcommand{\C}{\mathbb C}
 \newcommand{\DD}{\mathbb D}
 \newcommand{\eps}{\varepsilon}
 \newcommand{\To}{\longrightarrow}
 \newcommand{\BX}{\mathbf{B}(X)}
 \newcommand{\HH}{\mathfrak{H}}
 \newcommand{\D}{\mathcal{D}}
 \newcommand{\N}{\mathcal{N}}
 \newcommand{\W}{\mathcal{W}}
 \newcommand{\RR}{\mathbb{\mathcal}{R}}
 \newcommand{\HD}{\hat{\mathcal{H}}}
  \maketitle

 \begin{abstract}
        In this paper, using an exponential function of intensity of radiation field,
        two new classes of {\it nonlinear coherent states} will be constructed.
        For the first class, we choose the nonlinearity function as
        $f_{\beta}(n)=\exp(\beta n)$, where $\beta$ characterizes the strength of the nonlinearity
        of the quantum system.
        We show that, the corresponding $\beta$-states possess a
        collection of  non-classicality features,
        only for the particular values of $\beta$ and $z$.
        But, interestingly there exists  finite (threshold) values of $\beta$,  for which all of the non-classicality  signs will disappear,
        in appropriate regions  around the origin of the complex plane ($z < |Z|$).
        It is then illustrated that, using this threshold (or greater) value of $\beta$,
        the corresponding
        $\beta$-states behave very similar to {\it canonical coherent states}, as the most classical quantum states,
        in approximately whole of the space.
        In the continuation, we motivate to find  another class of nonlinear
        coherent states, limited to a unit disk centered at the origin, looking like the canonical
        coherent states in behavior, in exactly the whole range of $|z| <1$.
        This purpose also will be achieved by considering the nonlinearity function as
        $f_\lambda(n) =\exp(\lambda/n)/\sqrt n$, where $\lambda$
        is a tunable nonlinearity parameter.
        The canonical coherent state's aspects of the
        corresponding $\lambda$-states will be refreshed, in particular cases,
        working with a threshold (or greater)  value of $\lambda$.

 \end{abstract}


   \section{Introduction}\label{sec-intro}

    For a single mode radiation field, the {\it canonical coherent state} $|z\rangle_{ccs} $
    is obtained by the action of the displacement operator on the vacuum,
 \begin{equation}\label{displace}
   |z  \rangle _{ccs}= D(z)|0  \rangle=\exp(  z a^\dag - z^* a)|0  \rangle,
 \end{equation}
    or from the right eigenstate of annihilation operator
    $a|z\rangle _{ccs}= z |z\rangle _{ccs}$,
    where $z\in \mathbb{C}$ and  $a$, $a^\dag$ are  respectively the standard bosonic annihilation, creation
    operators with canonical commutation relation $[a, a^\dag]= \hat I$.
    The expansion of these states in the Fock space is written as follows:
 \begin{equation}\label{ccs}
   |z\rangle_{ccs} = e ^{-|z|^2/2}\sum_{n=0}^\infty
   \frac {z^n}{\sqrt {n!}}\;| n  \rangle, \qquad z\in \mathbb{C},
 \end{equation}
   where the set $\left\{|n\rangle\right\}_{n=0}^\infty$
   is the number states
   of the quantized harmonic  oscillator Hamiltonian
 \begin{equation}\label{hamilt}
    H=a^\dag a + \frac 1 2.
 \end{equation}
    With respect to these sets of states photon statistics is Poissonian  and the
    uncertainties in the two quadrature components are minimum, equal to the uncertainty of
    a vacuum of the radiation of field.

    Many attempts have been made on the generalization of these states to find the non-classicality  signatures.
    The motivation of the authors is mainly due to the
    increasing usefulness of coherent states in various branches of physics,
    from quantum optics (which the coherent states lie in it's core)
    and sensitive measurements, to foundations of quantum mechanics as well as
    other fields of physics \cite{application 1, application 2, application 3, agarwal_1, Schleish, Naderi, Bagerello}.
    In such a vast field of work, {\it "nonlinear"} or {\it "$f$-deformed"}
    coherent states as an algebraic generalization of coherent states
    have attracted much attention in recent
    two decades (for two most important pioneer papers see \cite{Matos1996, Manko1997}).
    Any class of nonlinear coherent states is characterized by a particular
    nonlinearity function $f(n)$. In this context, it has been proved that
    $q$-deformation can be identified in terms of $f$-nonlinearity by
    exponential functions as \cite{Manko1997}
  \begin{equation} \label{f-q}
     f_{q}(n)=\sqrt{\frac {\sinh{\gamma n}}{n\sinh{\gamma }}}=
     \sqrt{\frac {e^{\gamma n}-e^{-\gamma n}}{n
     ({e^\gamma-e^{-\gamma}})}},\qquad \gamma=\ln q,
  \end{equation}
    provided that $f_q(0)\doteq 1$,  $ 1 \leq q \leq \infty $ and $\gamma \in \mathbb{R}$.

    It has been recently shown that, nonlinear coherent states are useful in
    the description of the center of mass motion of a trapped ion \cite{Matos1996}.
    Various applications of these states can be found in the literature (as some
    interesting recent works see \cite{ bagheri, Honarasa}).
    There are so many generalized coherent states which
    can be classified  in this category with special nonlinearity functions
    \cite{Roknizadeh2004, Rokni-Tav_JMP}.
    Generally, unlike the canonical coherent states (\ref{ccs}), the nonlinear coherent states
    exhibit some fascinating  {\it "non-classical
    features"} as follows: (i) {\it quadrature squeezing} \cite{Walls}, (ii) {\it second order
    squeezing} \cite{Hillery, Hillery2}  (iii) {\it sub-Poissonian
    statistics} \cite{mandel, mandel2}   (iv) {\it antibunching effect} \cite{Glauber2},
    (v) {\it negativity of Wigner function in  parts of the phase
    space} \cite{wigner}
    (vi) {\it negativity of $A_3$ parameter} \cite{a3-agarwal}  and (vii)
    {\it oscillatory number distribution}.
    This collection of criteria, are widely used in the literature,
    discussing on the non-classicality of quantum states.
    Only one of the above properties is sufficient (not necessary) to establish the non-classicality of a
    state.

    Based on these frequently observations, the notion of "nonlinear coherent state"
    is generally considered to be synonym with the "non-classical states". This is
    not a strange conclusion, since to the best of our knowledge no exception has
    been reported up to now.
    Altogether,  a question may naturally be arisen:
    is there any class of nonlinear coherent states with a specific decomposition in
    the Fock space, i.e., a particular $f(n)$, which fails (or at least weaken) the latter conclusion?
    The main goal of the present contribution is to answer this question.
    Along achieving the purpose of paper,
    one should search for a "nonlinearity function" which the corresponding "coherent
    states" show neither of the usual non-classicality signs.
    As we will observe, in fact there exists such functions.
    In addition, we will illustrate that how the "nonlinear coherent states" associated  with the introduced nonlinearity
    functions, interestingly behave
    very similar to "canonical coherent states", as
    the particular states lie between "classical" and "non-classical" states.
    We first conjecture the appropriate case with considering the exponential nonlinearity
    function
  \begin{equation}\label{f-exp}
    f_{_{\beta}}(n)=e^{\beta n},
  \end{equation}
    where $\beta$ is an arbitrary tunable parameter, which characterizes the nonlinearity strength of the states.
    As we will demonstrate in the continuation (see definition (19) of the present
    paper),
    the corresponding nonlinear coherent
    states are defined in the whole space; i.e., $z \in \mathbb{C}$.
    Then,  a particular physical realization of the  nonlinear coherent
    states associated to (\ref{f-exp}), which has been called by us as $\beta$-nonlinear coherent
    states (or briefly $\beta$-states), will be deduced.
    Clearly $\beta = 0$ (or $f=1$) recovers the canonical coherent state in
    (\ref{ccs}),  which is known as "the most classical quantum state".
    At first sight on the nonlinearity function in (\ref{f-exp}), it may be expected that
    varying the value of $\beta$ from $0$ (the standard coherent states) results in growing up
    the nonlinearity strength and consequently the non-classicality exhibition of the $\beta$-states.
    But, while this will be occurred for some relatively low values of $\beta$, surprisingly according
    to our numerical results, this is not so for intermediate and large values of $\beta$(high nonlinearities).
    As we will show via the computational calculations,
    the behavior of the introduced $\beta-$ states
    is very close to the canonical coherent states in main features: Poissonian photon
    statistics, no antibunching and squeezing  exhibition (first and second order)
    for some critical values of $\beta$ in the relevant
    regions with $z > |Z|$. We continue the examination of  our results by evaluation of the Wigner
    function and  $A3$ parameter, too.
    We do not pay attention to the oscillatory number
    distribution of the introduced states, since it does not occur for them.
     Actually, there exists finite values of $\beta$ for which the
    $\beta$-states are canonical-like coherent states in a
    relatively wide interval of $|z|$, may be approximated by  the whole of the space.

    In the continuation of the paper, we motivate to search for a nonlinearity function whose the
    corresponding coherent states  show
    {\it "the most classical features"} in the whole range of the domain $z$. If we successfully
    find this, it  will be
    different essentially from previous $\beta$-states. As a matter of fact, we will observe that
    this purpose also will be achieved by
    considering the nonlinearity function as
  \begin{equation}\label{f-exp-1}
    f_{_{\lambda}}(n)=\frac {e^{\lambda /n}}{\sqrt n},
  \end{equation}
    where $\lambda$ is a tunable parameter.
    It must be noticed that, the corresponding nonlinear coherent states are restricted to a
    unit disk centered at the origin (see
    definition  (19) of the present paper).
    This fact makes a possible  opportunity for us to check the non-classicality signatures
    in the whole range of $|z| < 1$, numerically.
    One may recognize that, the special case $\lambda = 0$ (or $f(n)=1/\sqrt n$) in (\ref{f-exp-1})
    recovers the "harmonious states", previously introduced by Sudarshan in
    \cite{sudarshan},
    whose domain is limited to the unit disk around the origin. We
    will demonstrate that, all  of the mentioned criteria in the collection of
    non-classicality signs also disappear, for
    the  $\lambda$-states in the whole of the relevant space, when $\lambda$ becomes greater
    than a threshold value.

    At this stage it is worth noticing that the so-called {"Shr\"{o}dinger cat states"}
    are given by $\mathcal{N}(|z\rangle_{ccs} \pm |- z\rangle_{ccs})$ have Poissonian statistics and
    no squeezing when $|z| \gg 1$, meanwhile they show strong photon number oscillations.
    Also, "non-coherent states" recently introduced in \cite{Celia} have Poissonian
    statistics,  but they exhibit squeezing in one of the quadratures.
    As a result, both of the latter cases have been found in the
    earlier works are indeed non-classical states, since only one of the
    non-classicality signs may not be seen, while some others remain present.
    But, the main important feature of our $f$-deformed states, will be introduced in the
    paper, is that they show neither of the mentioned non-classical properties, when $\beta$
    or $\lambda$ are given the critical (or greater) values,  appropriately.
    In other words, considering the collection of mentioned non-classicality criteria,
    we proposed new classes of {\it "nonlinear coherent states"} possessing {\it "the most classical
    features"} and so {\it "the least nonclassical features"},
    at the level of "canonical coherent states".

   \section{Non-classicality criteria}

   For the manuscript to be self-contained,
   we briefly explain that some of the commonly used criteria in the
   literature, will be used by us for investigating the non-classicality exhibition of the states.
   Along this purpose, we refer to the sub-Poissonian
   statistics, antibunching phenomenon, quadrature squeezing and amplitude squared
   squeezing, negativity of Wigner function on phase space and finally negativity of $A_3$
   parameter.
   A common feature of all of the  above  criteria is that the corresponding Glauber Sudarshan
   $P$-function of a non-classical state is not positive definite.
   But, as a well-known fact, we would like to imply that finding this
   function is ordinarily a hard task.
   Altogether, each of the above effects, which will be considered in the paper, is indeed sufficient
   for a state to belong to non-classical states.
   It should be mentioned that, the necessary and sufficient criteria for the
   non-classicality of a state is the subject of recent researches \cite{Shcukin}.
 \begin{itemize}
 \item{\it Quadrature squeezing}\\
   Based on the dimensionless definitions of position and momentum
   operators,
   respectively as $x= (a+ a^\dag)/{\sqrt 2}$ and  $p= (a- a^\dag)/{\sqrt  2 i }$,
   the corresponding uncertainties will be defined as
   $(\Delta x)^2 \doteq  \langle x^2\rangle - \langle x\rangle ^2$ and
   $(\Delta p)^2 \doteq  \langle p^2\rangle - \langle p\rangle ^2$.
   A state is squeezed \cite{Walls} in $x$ ($p$) quadrature,
   if $S_x \doteq (\Delta x)^2 - 0.5 < 0$ ($S_p \doteq (\Delta p)^2 - 0.5 <0$).

 \item{\it Amplitude squared squeezing}\\
  In order to investigate amplitude-squared squeezing, the following two
  Hermitian operators have been introduced \cite{Hillery}
 \begin{equation}\label{X Y}
   X=\frac{a^2+{a^\dag}^2}{2},\qquad         Y=\frac{a^2-{a^\dag}^2}{2i}.
 \end{equation}
    In fact $X$ and $Y$ are the operators corresponding to the real and imaginary parts
    of the square of the complex amplitude of the electromagnetic field.
    The measurement of this effect through the interaction
    with a Kerr medium has been proposed in \cite{Hillery2}. Also, this property can be characterized by the methods
     proposed in \cite{Shcukin}.

   In the light of the Heisenberg uncertainty relation, the
   following amplitude-squared squeezing parameters may be defined,
 \begin{eqnarray}\label{I3 I4}
    I_{X} &=& \langle(\Delta X)^2\rangle - \frac{1}{2}\;|\langle [X,Y] \rangle|\nonumber\\
    &=& \frac{1}{4}\;\Big(\langle a^{4}\rangle+\langle{a^\dag}^{4}\rangle
    +\langle {a^\dag}^2 a^2\rangle+\langle a^2 {a^\dag}^2\rangle-
    {\langle a^2\rangle}^2-{\langle {a^\dag}^2\rangle}^2\nonumber\\
    &-& 2\;\langle a^2\rangle \;\langle {a^\dag}^2\rangle\Big)-\;\langle a^\dag a\rangle-\frac{1}{2},
 \end{eqnarray}
    and
 \begin{eqnarray}\label{I3 I4}
    I_{Y} &=& \langle(\Delta Y)^2\rangle  - \frac{1}{2}\;|\langle [X,Y] \rangle|\nonumber\\
    &=& \frac{1}{4}\;\Big(-\langle a^{4}\rangle-\langle{a^\dag}^{4}\rangle
    +\langle {a^\dag}^2 a^2\rangle+\langle a^2 {a^\dag}^2\rangle
    +{\langle a^2\rangle}^2+{\langle {a^\dag}^2\rangle}^2\nonumber\\
    &-& 2\;\langle a^2\rangle \;\langle {a^\dag}^2\rangle\Big)-\;\langle a^\dag a\rangle-\frac{1}{2}.
  \end{eqnarray}
   Therefore, negativity of $I_X$ ($I_Y$) indicates the amplitude squared squeezing
   in $X$ ($Y$) operators.
 %
   \item{\it Sub-Poissonian statistics}\\
    To examine the statistics of the states, Mandel's $Q$-parameter is often used,
    which is defined as \cite{mandel, mandel2}
 \begin{equation}\label{Mandel}
    Q= \frac{\langle n^2 \rangle - \langle n \rangle ^2}{\langle n \rangle} -1.
 \end{equation}
    This quantity vanishes for {\it "coherent
    light"}(Poissonian), is positive  for {\it "classical"} or {\it "chaotic light"}(supper-Poissonian),
    and negative
    for {\it "non-classical"} light(sub-Poissonian).
  \item{\it Second order correlation function}\\
  Even though there are  quantum states in which supper-/sub-Poissonian statistical
  behavior is together with bunching/antibunching effect, but this statement is not absolutely right.
  To investigate bunching or antibunching behavior, second order
  correlation function, is a useful quantity, defined as \cite{Glauber2}
 \begin{equation}\label{g2(0)2}
   g^{2}(0)=
   \frac{\langle{a^\dag}^2\;a^2\rangle}{\langle{a^\dag}\;a\rangle^2}.
 \end{equation}
   Depending on the particular nonlinearity function $f(n)$, has been chosen for the construction of nonlinear coherent states,
   $g^{2}(0)>1$ and  $g^{2}(0)<1$,
   respectively indicate  bunching and  antibunching effect. The case $g^{2}(0)=1$
   corresponds to the canonical coherent states.
   Note that, while for the Fock states with $n \geq 2$ one has $1 > g^2(0)=1-\frac 1
   n \geq 0.5$, zero value of this
   quantity is specified to  the states $|0\rangle,
   |1\rangle$ \cite{davidovich}. This is due to the fact that indeed,  for these two special
   number states the concept of antibunching and bunching
   seems to possess less meaning.
   So, upon the realization of the "vacuum" $|0\rangle$ as a "coherent state" $|z\rangle$  with
   $z=0$  (expanding the superposed states in (\ref{ccs}) and then setting $z=0$ in it),
   one may consider  two different values of the second order
   correlation functions for canonical coherent states, i.e.,
   zero for $z=0$, and $1$ for $0 \neq z \in \mathbb{C}$
   (recall that, the behavior of these two states, i.e.,
   vacuum and coherent states, are also exactly the same, when the variances in $x$ and
   $p$ are considered).

  \item{\it Wigner distribution function}\\
     As another indicator of non-classicality, one may refer to the
     negativity of Wigner distribution function \cite{wigner}. The Wigner function corresponding
     to any nonlinear coherent states obtained as follows \cite{Abbasi}:
  \begin{eqnarray}\label{wigner}
       W(z) &=& \frac{2}{\pi} (\mathcal{N}_f(|z|^2))^{-1} e^{ 2
       |z|^2}\sum_{m=0}^\infty \sum_{n=0}^\infty
       \frac{z^m
       (z^*)^n}{(-2)^{n+m}m![f(m)]!n![f(n)]!}\nonumber\\
       &\times& \frac{\partial^{n+m} }{\partial^n{(z^*)} \partial^m{z}}
       (-4|z|^2),
  \end{eqnarray}
     where $\mathcal{N}_f(|z|^2)$ is defined in (\ref{norm1}). Obviously, we will set $z=x+i p$ in the
     above relation in our next calculations.
     This function is positive in all space for canonical coherent states ($f(n)=1$).
     The negativity of this distribution function, which is a consequence of the
     selected nonlinearity function, can not be   interpreted classically
     and indicates that the state has no classical counterpart.
 \item{\it $A_3$ parameter}\\
   As another measure for testing the  non-classicality of any quantum state,
   Agarwal and Tara have introduced $A_3$ parameter
   \cite{a3-agarwal}.   Interestingly, the motivation of the authors  was
   searching a new criterion for non-classicality  of quantum states
   which does not show neither sub-Poissonian nor squeezing effects. They defined
  \begin{equation}\label{}
    A_3=\frac{\det m^{(3)}}{\det \mu^{(3)}-\det m^{(3)}}
  \end{equation}
  where
  \begin{equation}\label{}
     m^{(3)}= \left(
  \begin{array}{ccc}
    1 & m_1 & m_2 \\
    m_1 & m_2 & m_3 \\
    m_2& m_3 & m_4\\
    \end{array}
    \right),
  \end{equation}
   $m_n=\langle {a^\dag}^n a^n\rangle$ and $\mu^{(n)}$ can be
   obtained by replacing $m_n$  with $\mu_n=(a^\dag a)^n$.  Note
   that, specifically $\det (m^{(3)})=0$ for canonical coherent
   states and the vacuum. The lower bound of the normalized parameter
   $A_3$  is $-1$. So, it gets zero and $-1$ values, respectively for
   canonical coherent states and number states. Physically, the
   negativity of this parameter, known as a signature of the
   non-classicality of a state, comes out from phase-insensitive
   exhibition of the field state.

 \end{itemize}

 \section{Introducing a new class of nonlinear coherent states}

  The $f-$deformed ladder operators have been defined as \cite{Manko1997}
 \begin{equation}\label{f}
    A= af(n), \qquad A^\dag =f^\dag (n)a^\dag,
 \end{equation}
   where $n=a^\dag a$ and $f(n)$ is an "intensity dependent" function.
   Any class of nonlinear coherent state, characterizes by a function $f(n)$, is defined as the solution of the
   typical  eigenvalue  equation
   $A |z, f\rangle  = z |z, f\rangle $.  Their  decomposition in the number states basis read as
 \begin{equation}\label{NLCS}
   |z, f\rangle = \mathcal{N}_f(|z|^2)^{-\frac 1 2}\sum_{n=0}^\infty
   \frac {z^n}{\sqrt {n!}\;[f(n)]!}\;| n  \rangle,
 \end{equation}
   where  $[f(n)]! = f(n)f(n-1)\dots f(2)f(1)$ for $n>0$ and by convention $[f(0)]! \doteq 1$.
   The normalization
   constant  $\mathcal{N}_f(|z|^2)$ in (\ref{NLCS}) can  be readily calculated as
 \begin{equation}\label{norm1}
    \mathcal{N}_f(|z|^2) =  \sum_{n=0}^\infty \frac{{|z|}^{2n}}{n![f^\dag(n)]![f(n)]!}.
 \end{equation}
   Choosing different $f(n)$'s lead to distinct classes of nonlinear
   coherent states. Note that, we have assumed that the function  $f(n)$ appeared in the relations (\ref{f})-(\ref{norm1})
   to be generally a complex operator-valued function
   (for instance see  \cite {Rokni-Tav_JMP, Abbasi}).
   Anyway, these states are required to satisfy  the resolution of the identity as
 \begin{equation}\label{res}
    \int _D d^2 z |z, f\rangle W(|z|^2)\langle z, f | = \sum_{n=0}^\infty |n\rangle  \langle
    n|,
 \end{equation}
   where $d^2 z \doteq dx dy $.
   The nonorthogonality of a nonlinear coherent state, together
   with the the resolution of the identity requirement is known as the
   over-completeness relation \cite{Zubeiry}.
   The weight function $W(|z|^2)$ in (\ref{res}) is a positive definite function
   may be found after specifying $f(n)$, and $D$ is the domain of the states
   in the complex  plane determined by the disk
  \begin{equation}\label{disk}
    D = \{z \in \mathbb{C}, \; |z|^2  < \lim_{n\rightarrow \infty}
    n|f(n)|^2\},
  \end{equation}
   centered at the origin.

\subsection{$\beta$-nonlinear coherent states associated to $f_\beta( n)=\exp (\beta n)$}

     Along the aim of the paper, we now consider the nonlinearity function as in (\ref{f-exp}).
     It can be easily checked that, the following non-canonical commutation relation holds:
  \begin{equation}\label{fn}
     [A_\beta, A_\beta^\dag]= (n+1)e^{(n+1)\gamma} - n e^{n\gamma},
  \end{equation}
    where $A_\beta, A_\beta^\dag$ are respectively obtained from (\ref{f-exp}),
    (\ref{f}), and
    on the right hand side we have set $\gamma \equiv 2 \Re \; (\beta)$.
    Also, according to the nonlinear coherent states approach \cite{Manko1997}, the dynamics of the system is
    governed by the Hamiltonian $H_\beta(n)$, with the following action on the Fock space
  \begin{equation}\label{fn}
     H_\beta(n) | n \rangle  = \frac 1 2 [(n+1)e^{(n+1)\gamma} + n e^{n\gamma}] | n \rangle .
  \end{equation}
    Therefore, noticing that
  \begin{equation}\label{fact}
     [f_{_{\beta}}(n)]! = \prod _ {m=0}^{n} e ^ {\beta m} = e^{\beta n (n+1)/2
     },
  \end{equation}
   and upon using (\ref{NLCS}), one readily obtains the explicit decomposition of the $\beta$-states
   as follows
  \begin{equation}\label{NLCSf}
     |z, f_{_{\beta}}\rangle = \mathcal{N}_\beta (|z|^2)^{-\frac 1 2}\sum_{n=0}^\infty
     \frac {z^n e^{-\beta n (n+1)/2 }}{\sqrt {n!}}| n  \rangle,
  \end{equation}
     where the normalization factor is determined by
  \begin{equation}\label{normf}
     \mathcal{N}_\beta(|z|^2) = \sum_{n=0}^\infty
     \frac {|z|^{2 n} e^{-\gamma n (n+1)/2}}{n!},
  \end{equation}
     and as a general case $z, \beta \in \mathbb{C}$.
     Let us be more careful about the  values of $\beta$ and discuss the special cases which follow.
 \begin{itemize}

    \item
     The first case is $\beta \in \mathbb{C}$.
     From (\ref{disk}) it follows that $ D = \lim_{n\rightarrow \infty} n e^{\gamma n}$.
     Therefore, the associated states are defined on the whole of the complex plane
     if $\gamma \equiv 2 \Re (\beta) \geq 0$,
     independent of the imaginary part of $\beta$, otherwise the domain is $0$ and there is no well-defined states.
     The condition $\Re (\beta) \geq 0$ clearly illustrates why the so-called dual family of these states can not be
     defined in the Hilbert space \cite{RokniAli, royroy}.

   \item
     Secondly, we restrict $\beta$ to purely imaginary values.
     Replacing $\beta$ with $i \beta$ in (\ref{f-exp}) and limiting $\beta$ to real values, one obtains
     $f_{_{i \beta}}(n) = e^{i  \beta n}$
     and since the domain is given by $D = \lim n_{n \rightarrow \infty} n
     $,  one gets $z \in \mathbb{C}$, similar to the canonical coherent state.
     For this special case, the nonlinearity function  does not deform the canonical
     commutation relation, i.e., $[A_\beta, A_\beta^\dag]=\hat I=[a,
     a^\dag]$.
     The explicit expansion of the corresponding nonlinear coherent states then read as
  \begin{equation}\label{beta-imag}
      |z, f_{_{\beta}}\rangle = e^{-|z|^2/2}
      \sum_{n=0}^\infty
      \frac {z^n  e^ { -i \beta  n (n+1)/2 }} {\sqrt {n!}}\;| n
      \rangle.
  \end{equation}
      Hence, the photon-distribution of the obtained non-coherent
      states is exactly Poissonian, i.e., $P(n)= |\langle n |z, f_\beta \rangle|^2= \exp(-|z^{2} |)
      |z|^{2n}/n!$.
      From the group theoretical point of view, it can be easily observed that in this case the elements
      $\{A_\beta, A_\beta^\dag , A_\beta ^\dag A_\beta, \hat I\}$ constitute
      the generators of the Weyl-Heisenberg Lie algebra with closed commutation relations:
      $[A_\beta,\; A_\beta^\dag]=\hat I, [A_\beta,\;A_\beta^\dag A_\beta]=A_\beta,
      [A_\beta^\dag, \;A_\beta^\dag A_\beta]=-A_\beta^\dag$ .
      So, the $f$-deformed states in (\ref{beta-imag}) can be re-obtained through the
      action of the displacement operator
      $D(z)= \exp({z A_\beta ^\dag - z^* A_\beta})$ on the vacuum states $|0\rangle$. Recall that
      the nonlinear coherent states are not  generally of the displacement type states in
      exact form \cite{RokniAli, royroy}.
      Moreover, these states can be named as {\it "pseudo-canonical coherent states"}
      due to the above facts.
      Clearly, they are  not of the Gazeau-Klauder type \cite{{Gaz-Kl}}, especially due to
      the lack of the factor  $2^{-n}(n+1)!$ under the radical sign in the denominator of (\ref{beta-imag}).

  \end{itemize}

  \subsection{Production of the introduced states}

     Now, a brief view on the physical realization of a special case of the introduced $\beta$-states  will be offered.
     Keeping in mind that, $f(n) | m \rangle =f(a^\dag a) | m \rangle= f(m) |m\rangle$ (with $n=a^\dag a$ as the number operator),
     the $\beta$-states in (\ref{NLCSf}) can be rearranged via
  \begin{eqnarray}\label{realization1}
     |z, f_{_{\beta}}\rangle &=& \mathcal{N}_\beta(|z|^2)^{-\frac 1 2}
     e^ { - \beta n (n+1)/2 }
     \sum_{m=0}^\infty
     \frac {z^m }{\sqrt {m!}}\;| m  \rangle \nonumber \\
     &=& \mathcal{N}_\beta (|z|^2)^{-\frac 1 2} e^{|z|^2/2 + \beta/8} e^
     { -\beta H ^2 /2}| z \rangle_{ccs}, \qquad \beta \in
     \mathbb{C},
  \end{eqnarray}
     where $H$ is the Hamiltonian of the harmonic oscillator defined in (\ref{hamilt}).
     We now try to present a more realistic scheme to generate the states (\ref{beta-imag}).
     Following the same procedure led to (\ref{realization1}), the states
     in (\ref{beta-imag}) can be rewritten as
 \begin{eqnarray}\label{realization2}
     |z, f_{_{\beta}}\rangle &=& e^{-|z|^2/2}
     e^ { -i \beta  ({n}^2+ {n})/2 }
     \sum_{m=0}^\infty
     \frac {z^m }{\sqrt {m!}}| m \rangle \nonumber \\
     &=&  e^ { -i \beta  ({n}^2+  {n})/2 }  | z \rangle_{ccs},
     \qquad \beta \in \mathbb{R}.
  \end{eqnarray}
   Before we proceed, we should further emphasize on the fact that, the exponential functions of $n$,
   before sigma signs in the latter two relations, have the operational role
   (this situation is not more than the appearance of $n$ in (15), (20)-(22) and in the next sections of the paper, which $n$ has the
   "number operator" role, even though we have not used the notation $\hat n$).
   Anyway, apart from the coefficients, with the help of the derived result  in (\ref{realization2}),
   the $\beta$-states introduced in (\ref{beta-imag})  can be interpreted as the dynamical evolution of the standard
   coherent states (\ref{ccs}) traversing through nonlinear Kerr-like media,
   if the nonlinearity is considered to be $\chi(n)=n^2+n$.
   In other words, they can be generated by the action of time evolution
   operator $U(t)= \exp {(-i H t/\hbar)}$
   on the canonical coherent states, if $\beta$ parameter is replaced by  $2t$
   (after setting $\hbar=1=\omega$ ).
   This is actually the case that has been previously discussed in \cite{stoler, Kitagawa},
   leading to an interesting macroscopic superposition of distinguishable states.

 \subsection{The resolution of the identity of the $\beta$-states}

    In order to verify the resolution of the identity of the introduced
    $\beta$-states, which establishes the over-completeness of the states,
    we first insert (\ref{NLCSf}) in (\ref{res}) with $|z|^2 \equiv x$.
    Consequently, it can be easily seen that this property holds, if the following moment problem is satisfied
   \begin{equation}\label{res-fff}
     \pi \int _0^\infty dx \sigma (x,\gamma) x^n=
     n! e^{\gamma n (n+1)/2} ,\qquad n=0,1,2,\cdots,
   \end{equation}
    where $\sigma(x, \gamma) = \frac{W(x)} {\mathcal{N}_\gamma(x)}$.
    In fact, only a relatively small number of  $f(n)$'s are known for which the explicit
    form of  $\sigma$-functions can be extracted. Particularly,
    for purely imaginary value of $\beta$, i.e., $\gamma= 0$, equation (\ref{res-fff})
    has the simple solution $W(x)=1/\pi$.
    Moreover, since in the continuation of the paper we restrict ourselves to real $\beta$,
    to establish the resolution of the identity for $\beta \in \mathbb{R}$ we first put
   \begin{equation}\label{}
     \beta = \ln q, \qquad  1\leq q < \infty.
   \end{equation}
      The exponential nonlinearity function introduced
      in (\ref{f-exp}) takes then the simple form $f_{_{q}}(n)=q^n$.
      Therefore, the $\beta$-states obtained in (\ref{NLCSf}) can be expressed
      in terms of $q$-parameter rather than $\beta$ as follows
  \begin{equation}\label{nlcsq}
     |z, q\rangle = \mathcal{N}_q(|z|^2)^{-\frac 1 2} \sum_{n=0}^ \infty \frac {z^n}{\sqrt {n!}} q^{-n(n+1)/2}|
     n\rangle,
  \end{equation}
     with the normalization factor
  \begin{equation}\label{normq}
      \mathcal{N}_q(|z|^2)= \sum_{n=0}^\infty \frac {|z|^{2n}}{n!} q^{-n(n+1)}.
  \end{equation}
    Now, following the approach of Penson and Solomon in
    \cite{Penson},
    we can define a "generalized exponential function" $\varepsilon_q(z)$
    given by the differential equation
  \begin{equation}\label{}
    \frac {d \varepsilon_q(z)} {d z} = q^{-2} \varepsilon_q
    (q^{-2}z),
  \end{equation}
    where the series solution read as
  \begin{equation}\label{}
    \varepsilon_q(z)= \sum_{n=0}^\infty \frac {z^{n}}{n!} q^{-n(n+1)}.
  \end{equation}
    The case $q=1$ recovers the ordinary exponential function.
    Therefore, it is possible to reformulate the normalized states in (\ref{nlcsq}),
    using the above generalized exponential formalism
 \begin{equation}\label{}
    |z, q\rangle = \varepsilon_q(|z|^2)^{-\frac 1 2} \varepsilon_{q^{1/2}}(z a^\dag) |0\rangle\;.
 \end{equation}
     Also, the moment integral in (\ref{res-fff}) can now be expressed in terms of
     $q$-parameter as
  \begin{equation}\label{res-q}
    \pi \int _0 ^\infty  dx \sigma (x,q) x^n = n! q^{2 n (n+1)}.
  \end{equation}
      Based on the general theory of the Stieltjes power moment problem, a condition for solvability
      of  equation (\ref{res-q})
      is the positivity of the two series $\{h_0^{(n)}\}$ and $\{h_1^{(n)}\}$  known as
      Hankel-Hadamard matrices, defined by \cite{Akhizer}
  \begin{eqnarray}\label{}
   h_0^{(n)}(i,j)&=& (i+j-2)! q^{-2(i+j-2)(i+j-1)},\\ \nonumber
   h_1^{(n)}(i,j)&=&(i+j-1)! q^{-2(i+j-1)(i+j)} , \qquad n, i, j=1, 2, 3, \cdots \; .
  \end{eqnarray}
    It turns out that, indeed all the left-upper corner determinants of Hankel-Hadamard matrices are positive.
    So, the above discussion confirms us that a solution of equation (\ref{res-q})
    exists and therefore the resolution of the identity can be in fact satisfied.

    Altogether, we are now in a position to find the resolution of the identity in a more
    explicit fashion, using  Sixdeniers et al work in \cite{Sixdeniers}.
    Based on their obtained results, the Stieltjes moment  problem associated to the
    Mittag-Leffler generalized coherent states has the form
  \begin{equation}\label{res-mittag}
   \int _0 ^\infty  dx W (x;m,q) x^n = \frac{\Gamma({mn+1})}{q^{mn (n-1)}},
  \end{equation}
    where $n=0,1,2,\cdots$ and $m=1,2,3,\cdots$.
    After some lengthy procedure, they have finally deduced the following solution to the problem
  \begin{equation}\label{}
    W (x;m,q)  = \frac{1}{2m\sqrt{ \pi \delta} x}\int _0 ^\infty  du u^{\frac {1-m}{m}}
    e^{-u^{\frac{1}{m}} } \exp{\left(-\frac{(\ln(\frac{x}{q^mu}))^2}{4\delta}\right)},
  \end{equation}
   in which  $\delta \equiv m \ln {q^{-1}}$ and $0< q \leq 1$.
   Fortunately, the equation (\ref{res-mittag}) provides a good opportunity for us
   to solve the moment problem in (\ref{res-q}) and find $\sigma(x, q)$, straightforwardly.
   Indeed, this will be possible, provided that one sets $m=1$ with change of variables in
   the two following steps: firstly putting $x=q^2 y$ and then change $q$ by $q^{-2}$ in (\ref{res-mittag}).
   At last,  after some manipulations one gets the following result
  \begin{equation}\label{resq}
    \int _0 ^\infty  dy \left[ \frac{\sigma (y;1,q^{-2})}{q^4} \right ]y^n = n!q^{2n (n+1)}.
  \end{equation}
      with the solution
  \begin{equation}\label{res-qq}
     \sigma (x;1,q^{-2})= \frac{q^4}{2\sqrt{ \pi \delta} \;
     x}\int _0 ^\infty  du e^{-u}
     \exp{\left(-\frac{(\ln(\frac{x}{q^{-2}u}))^2}{4\delta}\right)},
  \end{equation}
     where $\delta =  \ln q^2 > 0$.
     So, the integral solution of (\ref{resq})  for $\sigma$ and therefore the principle
     solution of the moment problem in (\ref{res-q}) has been deduced.
     This conclusion has been  illustrated in figure (1) for some arbitrary values of $q$.
     It is worth also noticing that the other two principle requirements of the generalized coherent states
     $|z, q\rangle $ or equivalently $|z, f_\beta \rangle $, i.e., the continuity in the label $z$
     and the nonorthogonality
     of the states are so clear that need not to be investigated here.


   \subsection{Numerical results for non-classicality features of $\beta$-states}

   We are now ready to present our numerical results  taking into account the $\beta$-states in
   (\ref{NLCSf}), which will demonstrate the aim of our contribution.
   In figures  2 and 3  we have plotted $S_x$ and $S_p$,  respectively
   as a function of
   $\beta \in \mathbb{R}$ for different fixed values of $z\in \mathbb{R}$, and the
   same parameters as a function of  $z\in \mathbb{R}$
   for different fixed values of $\beta \in \mathbb{R}$,
   utilizing  the $\beta$-states.
   It is seen that, squeezing occurs in $x$ quadrature only up to $\beta \simeq
   5.5$,
   for the selected values of $z$ have been used in numerical calculations.
   But, increasing $z$ causes an increase in  $\beta$, for which quadrature squeezing disappear.
   Figures 4 and 5 deal with the amplitude squared squeezing  against
   $\beta$,
   for some fixed values of $z$, and the same parameter  against  $z$  for different fixed values
   of $\beta$, respectively.
   As it may be observed, the amplitude squared squeezing occurs in $X$, and both of $I_X$, $I_Y$
   will be vanished for $\beta > 2$.
   In figures 6 and 7, the Mandel's $Q$-parameter has been
   shown as a function of
   $\beta$ for various fixed $z$, and  against $z$ for various values of $\beta$,   respectively.
   Sub-Poissonian statistics is observed from figure 6, in finite intervals of $\beta$ for fixed values of
   $z$. Also, from figure 7 this non-classicality sign is seen for all $z$ considered,
   when some particular values of $\beta$ have been used.
   With increasing $z$, the negativity of $Q$-parameter and the range of it will become deeper and wider.
   But, this phenomena will be disappeared for $\beta \geq 7.5$ in the range of $z$ that was used in obtaining figure 7.
   Consequently,  for  $\beta  \geq 7.5$ the
   Mandel's $Q$-parameter tends to zero, indicating the canonical coherent state behavior  for the considered values of $z$.
   Figure 8 shows the second order correlation factor for
   $\beta$-states. In all of the graphs displayed in Fig. 8, which have been plotted for different
   fixed values of $z$, the values of $g^2(0)$ begin from $1$ at
   $\beta=0$ ($f(n)=1$), as it may be expected. But, with increasing $\beta$, this quantity
   becomes less than $1$ (showing antibunching effect) and finally decrease to zero, irrespective of the value of $z$.
   So, clearly there exists finite values of $\beta$, for which this function vanishes.
   To explain the obtained results along the aim of the present paper,
   recall that for the vacuum, one has $g^2(0)=0$.
   Therefore, according to the results displayed in Fig. 8  for which $g^2(0)$
   vanishes for  $\beta >3$, one may conclude that at this range of $\beta$, the $\beta$-states behave like vacuum
   (as a coherent state with $z=0$).
    We also continue our investigation on the non-classicality of $\beta$-states
    by studying the Wigner distribution function presented  in Fig. 9.
    We have used $z=200$ in all cases, but in figures 9-a,
    9-b, 9-c we have respectively set $\beta=2,  4, 7.5$. From the figures it is obviously seen that,
    with increasing $\beta$, the regions and depth of the negativity of Wigner function
    on the phase space will be decreased. When the threshold value of $\beta \simeq 7.5$ is chosen, this non-classicality feature disappears.
    Our calculation on $A_3$ parameter has been shown in figure 10.
    From the plotted graphs it is clear that, all of them which have
    been plotted for the different values  of $z$, begin from zero,  become negative and finally tend
    to zero (the specific value of $A_3$ parameter for canonical coherent states).
    So, by this criteria, non-classicality sign is visible for all cases represented in the figures, up to a particular value of $\beta$.
    But, it is found that increasing $\beta$ from $\simeq 2.5$, vanishes
    this quantity for all chosen values of $z$, i.e., this criteria is not as strong as some of the previously outlined
    ones to test the non-classicality of this set of states. It
    must be noted that in plotting figures 9 and
    10, we have used $z=200$, to show that how the non-classicality features behave at this value.
    Obviously, decreasing the value of $z$ down to $\simeq 15$,
    which has been used in the previous figures, decreases the
    threshold value of $\beta$ for these two criteria. For
    instance, the negativity of Wigner function will be lost at $\beta\simeq
    3$, when $z=15$ is used.


   Adding the above results, have been represented in figures 2-10, related to
   $\beta$-states, showing that, generally an increase in $\beta$ from
   zero (canonical-like coherent state's behavior), firstly followed by a sudden increase in the degree of
   non-classicality, but all of the signs will be disappeared gradually,
   for  the threshold (or greater) values of $\beta$ and chosen values of
   $z$.

   Altogether, keeping in mind the obtained results, we are not yet at a point to conclude the goal of the paper,
   due to the sensitive dependence of the studied effects on both $z$ and $\beta$, simultaneously.
   Nevertheless, our further computational evaluations (with precession $\pm 0.01$) show that:
   (i)
   for each $\beta$ there exists a disk  with finite radius $|Z|$
   around the origin of complex plane, such that
   all of the non-classical features will be vanished (one may call $|Z|$ as the "{\it radius of coherence}"),
   (ii)
   with smoothly increasing $\beta$,
   the value of $|Z|$ will be  considerably increased.
   As some specific examples, while the radius of coherence is $|Z|=15$ for $\beta =5$, this becomes $|Z|=200$
   for $\beta=7.5$, $|Z|=5 \times 10^7$
   for $\beta=20$, and $|Z|=10^{12}$ for $\beta=30$. So, a low increase in $\beta$,
   unexpectedly makes the radius of coherence $|Z|$ wider and wider.
   Consequently, it is not far from the fact, if one concludes that there exists a special set of
   $\beta$-states with intermediate
   finite values of $\beta$, which are "canonical-like coherent states" in a relatively wide
   ranges of $z$ defined by $|Z|$.
   It may be clear that, the radius of coherence can be enlarged arbitrarily,
   using  the value of $\beta$, properly. Therefore, there exists some
   $\beta$-states, so that their behavior become very similar to "the most classical quantum states" in
   approximately the whole of the space, i.e., as the case of
   canonical coherent states.

 \section{Another class of nonlinear coherent states with least nonclassical properties}

  Now,  we motivate to search for a nonlinearity function whose corresponding coherent states
  are limited to the unit disk centered at the origin, showing {"the most classical features"} and so "the least nonclassical
  properties",in the whole range of $|z| < 1$.
  If there exists such a function, the associated coherent states  will be
  apparently different from the previously introduced case in (\ref{NLCSf}).

 \subsection{$\lambda$-nonlinear coherent states associated to $f_{_{\lambda}}(n)={\exp(\lambda /n)}/{\sqrt n}$}

  Specifically, as we will observe, the demand outlined in this section will be achieved by considering the
  $\lambda$-dependent nonlinearity function, suggested by us in (\ref{f-exp-1}).
  A few words seems to be necessary in relation to the introduced function.
  At first glance, it may be seemed that for the special case of $n=0$,
  the $\lambda$-nonlinearity in  (\ref{f-exp-1}) is ill-defined.
  But, keeping in mind the explanations which we presented
  after equation (\ref{NLCS}) based on the general formalism
  of nonlinear coherent states method  and  recalling
  some of the previous well-known
  nonlinearity functions such as harmonious states with $f(n)=1/\sqrt n$
  introduced by Sudarshan \cite{sudarshan} and $q$-deformed coherent states which are
  well established nonlinear coherent states with $f_q(n)$ introduced in
  (\ref{f-q}), can solve the ambiguity. Indeed, generally in the "nonlinear coherent states
  method" we deal with   $[f(0)]!$, not $f(0)$  itself (see the expansion (\ref{NLCS})), which conventionally
  assumed to be $1$ \cite{Manko1997, Manko2}.  Anyway, the actions of $\lambda$-deformed ladder operators
  (obtained from (\ref{f-exp-1}) and (\ref{f}))
  on the number states may be expressed as
  \begin{eqnarray}\label{}
     A_\lambda^\dag |n \rangle &=& \exp{[\lambda/(n+1)]} |n +1
     \rangle, \qquad n \geq 0, \nonumber \\
     A_\lambda |n \rangle &=& \exp{(\lambda/n)} |n -1 \rangle, \qquad  n
     \geq  1,
 \end{eqnarray}
    and by definition $A_\lambda |0 \rangle \doteq 0$. Also, they obey the following commutation relation
    with the action
 \begin{equation}\label{}
      [A_\lambda, A_\lambda^\dag] | n \rangle= \left(\exp(2 \lambda/(n+1)) - \exp(2
      \lambda/n)\right) | n \rangle,  \qquad n\neq 0,
 \end{equation}
   while for $n=0$ one has $[A_\lambda, A^\dag _\lambda] | 0 \rangle = \exp(2 \lambda) | 0
  \rangle$.    The dynamics of the system, is governed by the Hamiltonian whose action may be expressed as  \cite{Manko1997}
 \begin{eqnarray}\label{fn}
     H_\lambda(n) | n \rangle &=& \frac 1 2 [\exp(2 \lambda/(n+1)) + \exp(2 \lambda/n)] | n \rangle , \qquad n\neq 0
     \nonumber\\
     H_\lambda(n)| 0 \rangle &=& \frac 1 2 [\exp(2 \lambda)] | 0 \rangle, \qquad n=0.
  \end{eqnarray}
   To this end, the nonlinear coherent states for the introduced  $\lambda$-nonlinearity read as
 \begin{equation}\label{NLCSf-1}
   |z, f_{_{\lambda}}\rangle = \mathcal{N}_{\lambda}(|z|^2)^{-\frac 1 2}\sum_{n=0}^\infty
   \frac {z^n}{[e^{\lambda /n}]!} |n \rangle,
 \end{equation}
    where the normalization factor is determined by
 \begin{equation}\label{normf}
     \mathcal{N}_{\lambda}(|z|^2) = \sum_{n=0}^\infty
     \frac {|z|^{2 n}} {([e^{\lambda /n}]!)^2}.
 \end{equation}
    Note that, using the $\lambda$ nonlinearity of (\ref{f-exp-1}) in (\ref{disk}) makes it clear that the states
    in (\ref{NLCSf-1}) can be defined  only in $|z|<1$.
    Obviously, the $\lambda$-states are different from the previously introduced $\beta$-states in (\ref{NLCSf}).

 \subsection{Numerical results for non-classicality features of $\lambda$-states}

   Our results, utilizing the $\lambda$-states  in (\ref{NLCSf-1}), will be presented in this subsection.
   We confine ourselves to real and positive values of $\lambda$ to
   achieve the purpose of paper.
   In figures 11 and 12, the squeezing parameters, respectively in $x$ and $p$, have been
   plotted as a function of $\lambda$, for different fixed values of
   $z$,  and  as a function of $z$ for various fixed values of
   $\lambda$.
   As it is observed, quadrature squeezing occurs in $p$ for all values of $z$, only when particular finite values of $\lambda$ have been used.
   In figures 13 and 14 we have displayed the amplitude squared squeezing
   against $\lambda$  for some choices of $z$,
   and as a function of $\lambda$ for some fixed values of $z$, respectively.
   As it is seen, the amplitude squared squeezing in
   $Y$ has been  occurred for some used values of $\lambda$.
   A common feature of the figures 11-14 is that, the squeezing
   (first and second orders) completely disappears  for intermediate $\lambda$, say $\lambda \geq 5.25$.
    Following the purpose of the paper, we have plotted Mandel's $Q$-parameter,  respectively in figures 15 and 16 against $\lambda$,
    for different values of $z$, and against $z$
    for different values of $\lambda$.
    From the figures it is obvious that, the photon statistics is super-Poissonian and
    increasing $\lambda$ tends the graphs to the horizontal line at  $Q=0$ for $\lambda \geq 5.25$.
   Figure 17 shows the second order correlation function for
   $\lambda$-states. From the figure it is observed that, this
   function begins from $2$, and becomes greater and greater as $z$ increases (bunching effect is visible).
   Therefore, these states do not possess this non-classicality sign at all (they have bunching behavior).
    For further investigation of the non-classicality of $\lambda$-states, the variation of
    Wigner distribution function on phase space has been employed. The results are displayed in figure 18.
    In our calculations
    we have used $z=0.95$ in all cases, but in 18-a,
    18-b, 18-c we have respectively set $\lambda=0$ (harmonious
    states  \cite{sudarshan}), $0.5, 2$. From the plotted graphs  it is clear that increasing
    $\lambda$ from zero,  decreases the regions of the negativity of the Wigner function
    on the phase space. This non-classicality signature also disappears,
    even at value of $\beta$ less than those we have indicated for previously considered criteria.
    Finally, our calculation on $A_3$ parameter has been shown in figure 19.
    From the figure it is observed that, in all cases, the $A_3$ parameter begins from a positive value,
    gradually grow up with increasing $\lambda$, gets some peaks and
    gradually reduces to zero, i.e., the specific value of $A_3$ parameter for canonical coherent states.
    So, generally the $\lambda$-states may not possess this non-classicality sign, at
    all. While, for $\lambda \geq 3.5$, $A_3$ parameter does not recognize
    between $\lambda$-states and canonical coherent states.

   Consequently, adding the numerical results in figures 11-19, the purpose of the paper,  is
   achieved. In fact,
   the behavior of the introduced $\lambda$-states in (\ref{NLCSf-1}) are very similar
   to the canonical coherent states,
   for some relatively small values of $\lambda \simeq 5.25$ and certainly greater than this threshold value.
   It is remarkable that, since the $\lambda$-states are defined only on $|z|<1$,
   one can be sure that, there is no possibility for either of the non-classicality signs, to arise,
   again, when the threshold value of
   $\lambda$ has been chosen, properly.

 \section{Summary and conclusion}

    In summary, we have first introduced a set of $f_\beta$-deformed coherent states with the special
    nonlinearity function
    $f_\beta(n) = \exp(\beta n)$. Generation a special case of the $\beta$-states through some simple
    nonlinear Kerr-like media
    is then illustrated. It is shown that generally these states possess some of the routine non-classical
    properties such as first and second order squeezing, sub-Poissonian statistics, antibunching,  negativity of
    both Wigner function and $A3$ parameter. Indeed, increasing the value of $\beta\in \mathbb{R}$ from zero
     causes suddenly increase in the degree
    of non-classicality, but it gradually tends to zero, i.e., all of the considered non-classicality signs will be
    disappeared for a threshold value of (or greater than) $\beta$.
    Strictly speaking, for each specific choice of  $\beta$, there exists certain  interval of $|Z|$ around the
    origin of space (has been called by us as the radius of coherence) for which $\beta$-states behaves like
    canonical coherent states. But, fortunately this
    interval can be arbitrarily enlarged using the value
    of $\beta$, properly.
    Thus, in this sense, while the explicit representation of $\beta$-deformed states
    and canonical coherent states in the Fock space are apparently different,
    they are not distinguishable from the point of view of "non-classical" behavior.
    Along the goal of the paper, we motivated to search for another nonlinearity function which the
    corresponding coherent states show "the most
    classical features" in exactly whole of the relevant space. We observed that these states also may exist with
    $f_{_{\lambda}}(n)=\frac {e^{\lambda /n}}{\sqrt n}$, whose the associated coherent states are limited to $|z|<1$.
    In fact, there exists certain values of
    $\lambda \geq 5.25$, for which all of the non-classicality features will be  disappeared in the whole range of $|z|<1$.

    It should be noticed that, the threshold values of $\beta$ or $\lambda$,  for each of
    the non-classicality criteria, in the corresponding states are not exactly the same.
    Moreover, the point is that there is proper finite values of
    them for which  all of the considered non-classicality signs failed to be observed.
    Therefore, we were successful in finding two sets of nonlinear coherent
    states, whose behavior are totally,  very similar to canonical coherent states.
    A deep insight in the details of our formalism and the represented graphs, may lead one to conclude that
    in working with the two classes of introduced states,
    using their threshold (or larger) values of corresponding $\beta$ and $\lambda$,
    the ground (vacuum) state has dominant contribution in the superposed states, relative to other
    number states. Altogether, all of the number states
    are clearly present in each state.

   Summing up, we conclude the paper with the following  results;
   (i) It is not far from the fact, if the introduced $\beta$- and
   $\lambda$-states are also known as
   quantum states with nearly the most classical aspects.
   (ii) "Nonlinear coherent states" do not necessarily
   possess "non-classicality  features".
   (iii) While the  nonlinearity causes non-classicality of states, one may not
   always yield an increase in the non-classicality features with growing  the
   strength of what is called the nonlinearity function.
   (iv) An analysis of our results represented in the numerous figures shows that,
   quadrature squeezing and Mandel parameter, at least, among the collection of non-classicality criteria and
   associated with the two classes of states have been considered by us, seems to be more sensitive
   for examining the non-classicality behavior,
   since they lead to some greater upper bounds of $z$  to exhibit  non-classicality.
   At the same time, $A_3$ parameter behaves as the weakest criteria.

    Nevertheless, it is worth to mention that, unlike the canonical coherent states,
    neither of the introduced nonlinear coherent
    states possesses the temporal stability property. But, this does not have any effect on
    our discussion on the non-classicality of states.
    At last, we end the summary with pointing the
    important advantage of the two introduced sets of "nonlinear coherent
    states" as compared with "canonical coherent states" in computational
    manipulations. For computing the non-classical effects of the states have been discussed in the
    present paper, the involved series are converged more rapidly and so the results are very
    certain.
    This is while, for the same calculations regarding canonical coherent states, it
    is necessary to take into account a lot of extra
    terms in the involved series, to make certainty about the convergence and to obtain exact results.

 {\bf Acknowledgements:} We are thankful to the referees for their
                        valuable comments and suggestions, which improved the clarity and
                        enriched the contents of the paper. Also,
                        thanks to the Research Council of Yazd
                        University for their financial supports of
                        this project.

 \include{thebibliography}

 \newpage
     {\bf FIGURE CAPTIONS}
   \vspace {1.5 cm}

   {\bf FIG. 1}   The graph of weight function $\sigma(x;1,q^{-2})$ in (31) as a function of
                  $x$,
                  for different values of $q$. Continuous line is plotted for $q=2$, dotted for $q=5$
                  and dot-dashed for $q=10$.

   \vspace {.5 cm}

  {\bf FIG. 2} The graph of squeezing parameters in $x$ and $p$ against $\beta \in \mathbb{R}$, for
               different values of $z$, utilizing  the
               $\beta$-states were  introduced in (\ref{NLCSf}).
               Dotted lines are plotted for
               $z=1$, dot-dashed for $z=2.5$, tiny dashing for $z=5$, large dashing for $z=10$ and continuous
               lines for $z=15$ (curves with negative values correspond to $s_x$).

  \vspace {.5 cm}

  {\bf FIG. 3}  The same as figure 2
                except that it is plotted against $z \in \mathbb{R}$, for different fixed values of $\beta$.
                Dotted lines are for $\beta=0.5$, dot-dashed for $\beta=1$, tiny dashing for $\beta=2.5$,
                large dashing for $\beta=5$ and continuous lines which coincides with the horizontal axis are for $\beta=7.5$
                (curves with negative values correspond to $s_x$).

  \vspace {.5 cm}

 {\bf FIG. 4} The graph of amplitude squared squeezing effects in $X$ and $Y$ ($I_X, I_Y$) against
              $\beta \in \mathbb{R}$, for
              different values of $z$,
              utilizing  the $\beta$-states were  introduced in (\ref{NLCSf}). Continuous  lines are plotted for
              $z=2.5$, dotted for $z=5$, dot-dashed  for $z=10$ and  dashed for
              $z=15$. Amplitude squared squeezing may be occurred in the $X$
              component.
  \vspace {.5 cm}

  {\bf FIG. 5} The same as figure 4
               except that it is plotted against $z \in \mathbb{R}$, for different fixed values of $\beta$.
               Continuous lines are for $\beta=0.25$, dotted  for $\beta=0.5$, dot-dashed for
               $\beta=1.5$ and dashed line which coincides with the horizontal
               axis is for $\beta=2$. Amplitude squared squeezing may be occurred in the $X$
               component.

  \vspace {.5 cm}

  {\bf FIG. 6} The graphs of Mandel parameter as  functions of
               $\beta \in \mathbb{R}$, for
               different values of $z$, are
               utilizing the $\beta$-states in (\ref{NLCSf}).
               Dotted line is for $z=1$, dot-dashed for $z=2$, tiny dashing for $z=5$, large dashing
               for $z=10$ and continuous line for $z=20$.

  {\bf FIG. 7} The same as figure 6
               except that it is plotted against $z\in \mathbb{R}$, for different values of $\beta$.
               Dotted line is for $\beta =0.5$, dot-dashed for $\beta=1$, tiny dashing for $\beta=2.5$,
               large dashing for $\beta=5$ and continuous thick line for $\beta=7.5$ (coincides with the
                horizontal axis).

  \vspace {.5 cm}
  {\bf FIG. 8}
               The graphs of second
               order correlation function
               against $\beta  \in \mathbb{R}$, for
               different values of $z$,
               utilizing the $\beta$-states in (\ref{NLCSf}).
               Dotted line is for $z=1$, dot-dashed for $z=2$, tiny dashing for $z=5$, large dashing
               for $z=10$ and continuous line for $z=20$.

 \vspace {.5 cm}

 {\bf FIG. 9}   The graph of Wigner function on phase space for
               $\beta$-states introduced in (\ref{NLCSf}). We have used $z=200$ in all cases, but in 9-a, 9-b,
               9-c we have respectively set $\beta=2, 4, 7.5$.

 \vspace {.5 cm}

 {\bf FIG. 10} The graph of $A_3$ parameter for
              $\beta$-states introduced in (\ref{NLCSf}).  Continuous line is for $z=5$, dotted line for $z=10$,
              dot-dashed line for $z=15$, and large dashing line for $z=200$.
 \vspace {.5 cm}

 {\bf FIG. 11}  The graph of squeezing parameters in $x$ and $p$
              against $\lambda \in \mathbb{R}$, for
              different values of $z$,
              utilizing  the $\lambda$-states were  introduced in (\ref{NLCSf-1}).
              Dotted lines are plotted for
              $z=0.25$, dot-dashed for $z=0.5$, continuous line for
              $z=0.75$ and large dashing for $z=0.9$ (the negative graphs correspond to $s_p$).

 \vspace {.5 cm}

 {\bf FIG. 12} The same as figure 11
              except that it is plotted against $z \in \mathbb{R}$, for different fixed values of $\lambda$.
              Dotted lines are for $\lambda=0.5$, dot-dashed for $\lambda=1$, tiny dashing for $\lambda=2.5$,
              large dashing for $\lambda=5$ and continuous lines which coincides with the horizontal axis are for
              $\lambda=5.5$ (the negative graphs correspond to $s_p$).

 \vspace {.5 cm}

 {\bf FIG. 13} The graph of amplitude squared squeezing
              effects in $X$ and $Y$ ($I_X, I_Y$) against $z \in \mathbb{R}$, for
              different values of $\lambda$,
              utilizing  the $\lambda$-states were  introduced in (\ref{NLCSf-1}). Dotted lines are plotted for
              $z=0.25$, dot-dashed  for $z=0.5$, tiny dashing for $z=0.75$  and  large dashing for
              $z=0.9$ (the negative graphs correspond to $I_Y$).

 \vspace {.5 cm}

 {\bf FIG. 14} The same as figure 13
              except that it is plotted against $z \in \mathbb{R}$, for different fixed values of $\lambda$.
              Continuous lines are for $\lambda=0$, dot-dashed for $\lambda=1$, tiny dashing for
              $\lambda=2$ and large dashing which coincides with the horizontal axis is for $\lambda=4$
              (the negative graphs correspond to $I_Y$).

 \vspace {.5 cm}

 {\bf FIG. 15} The graph of Mandel's $Q$-parameter as a
             function of  $\lambda \in \mathbb{R}$,
             for different fixed values of $\lambda$,   using the introduced $\lambda$-states in (\ref{NLCSf-1}).
             Dotted line is for $z=0.25$ (harmonious states), dot-dashed for $z=0.5$,
             large dashing for $z=0.75$ and continuous
             lines  for $z=0.9$.

 \vspace {.5 cm}

  {\bf FIG. 16} The same as figure 15
                except that it is plotted against $z\in \mathbb{R}$, for different values of $\lambda$.
                Dotted line is for $\lambda=0$ (harmonious states), dot-dashed for
                $\lambda=1$, tiny dashing for $\lambda=2.5$, large dashing
                for $\lambda=5$ and continuous lines for $\lambda=5.5$ (coincides with the horizontal axis).

 \vspace {.5 cm}

  {\bf FIG. 17} The graph of second order correlation function against
                $\lambda \in \mathbb{R}$, for different fixed values of $z$, using the introduced $\lambda$-states in (\ref{NLCSf-1}).
                Dotted line is for $z=0.25$, dot-dashed for $z=0.5$,
                tiny dashing for $z=0.75$ and continuous
                lines  for $z=0.95$.

 \vspace {.5 cm}

 {\bf FIG. 18}  The graphs of Wigner function on phase space for
                $\lambda$-states introduced in (\ref{NLCSf-1}). We have used z=0.95 in all cases, but in 18-a,
                18-b, 18-c we have respectively set $\lambda=0$ (harmonious
                states), $0.5, 2$.

 \vspace {.5 cm}

 {\bf FIG. 19} The graph of $A_3$ parameter against $\lambda\in
               \mathbb{R}$, for
              $\lambda$-states introduced in (\ref{NLCSf-1}).  Dot-dashed line is for $z=0.5$, dashing line is for $z=0.75$,  continuous
              line for $z=0.85$, dotted line is for $z=0.95$ and continues thick line for
              $z=0.99$.

 \end{document}